# Attainment of Gigavolt Potentials by Fluid Dynamic Suppression of the Stepped Leader – its Significance for Thermonuclear Ignition

F. Winterberg
University of Nevada
Reno

**Abstract**

It is proposed to levitate a conducting sphere in a high pressure Taylor flow and to charge it up to gigavolt potentials, either mechanically as in a Van de Graaff electrostatic generator, or inductively by a rising magnetic field. If the Taylor flow is sufficiently fast, it should overcome the electric pressure and breakdown by stepped leader formation, leading to the maximum attainable voltage by the Paschen law. Discharging the electrostatically stored energy can be done by controlled breakdown. With gigajoule energies stored and released in about $10^{-8}$ sec, this implies and electric pulse power of the order $10^{17}$ Watt, opening the prospect of large driver energies for thermonuclear ignition.

# 1. Introduction

A typical lightning bolt has an energy of several hundred megajoule, discharging several coulombs with a current of 10 – 100 kA. Lightning occurs if the electric field between a cloud and the ground exceeds the breakdown of air at about 30 kV/cm. For a 300 meter long lightning this implies a potential difference of $10^9$ Volt, and for a current of 100 kA a power of $10^{14}$ Watt. Most lightning discharges are from a negatively charged cloud to the ground, but at rare occasions from a positively charged cloud to the ground. In the rare cases the current may reach 300 kA discharging 300 Coulombs. At the potential difference of $10^9$ Volt an energy of 300 gigajoule is released. This is equal to the energy released by 75 tons of TNT, with a power of $3 \times 10^{14}$ Watt. By comparison, to ignite in liquid deuterium-tritium a thermonuclear micro-explosion, requires an energy of about 10 MJ with a power of $10^{15}$ Watt. This raises the question if one cannot make artificial lightning, comparable in energy and power of natural lightning, and use the thusly released energy to drive inertially confined thermonuclear micro-explosions. One way it conceivably can be done is by charging to gigavolt potentials a magnetically insulated conductor levitated in ultrahigh vacuum (Winterberg, 1968, Winterberg, 2000). Here I will describe a scheme by which the same goal can be reached in a very different way.

The reaching out for high voltages is of importance in the quest for the ignition of thermonuclear reactions for two reasons:

1. With the energy E [erg] stored in a capacitor C [cm] charged to the voltage V [esu] equal to

$$E = (1/2) CV^2, \qquad (1)$$

having an energy density

$$\varepsilon \sim E/C^3 \sim V^2/C^2, \qquad (2)$$

the energy is discharged in the time τ [sec]

$$\tau \sim C/c, \qquad (3)$$

with a power P [erg/s]

$$P \sim E/\tau \sim cV^2 \qquad (4)$$

This shows that for a given dimension of the capacitor measured in its length, and hence volume, the energy stored and power released goes in proportion to the square of the voltage.

2. If the energy stored in the capacitor is released into the energy of a charged particle beam, the current should be below the critical Alfven limit

$$I = \gamma \, I_A \quad (5)$$

where $\gamma = (1-v^2/c^2)^{-1/2}$ is the Lorentz boost factor, and $I_A = mc^3/e$. For electrons $I_A = 17$ kA, but for protons it is 31 MA. Only for $I \ll \gamma I_A$, can one view the beam as a beam of particles accompanied by the field of the particles, while for $I \gg \gamma I_A$ it is better viewed as an electromagnetic pulse carrying with it along some particles. For $I \gg \gamma I_A$, the beam can propagate in a space-charge and current-neutralizing plasma, but only if $I \leq \gamma I_A$ can the beam be easily focused onto a small area, needed to reach a high power flux density. If a power of $\sim 10^{15}$ Watt shall be reached with a relativistic electron beam produced by a $10^7$ Volt Marx generator, the beam current would have to be $10^8$ Ampere with $\gamma \cong 20$ and $\gamma I_A \sim 3 \times 10^4$ Ampere, hence $I \gg \gamma I_A$. But if the potential is $10^9$ Volt, a proton beam accelerated to this voltage and with a current of $I = 10^7$A is well below the Alfven current limit for protons. And it would have the power of $10^{16}$ Watt, sufficiently large to ignite the D-D thermonuclear reaction.

## 2. Electric Breakdown below the Paschen Limit as an Electrostatic Instability

According to Paschen's law the breakdown voltage between two plane parallel conductors is only a function of the product pd, where p is the gas pressure and d the distance between the conductors. For dry air the breakdown voltage is $3 \times 10^6$ V/m, such that for a distance of 300m the breakdown voltage would be $10^9$ Volt. This is the voltage which under ideal condition is reached in a lightning discharge. In reality the breakdown voltage is much smaller. The reason is that by a small initial inhomogeneity in the electric field, more negative charge is accumulated within the inhomogeneity, further increasing the inhomogeneity and eventually forming a "leader", a small luminous discharge of electrons bridging part of the distance between the electrodes with a large potential difference. As a result a much larger electric field inhomogeneity is created at the head of the "leader", which upon repetition of the same process leads to a second "leader", followed by a third "leader", and so on, resulting in a breakdown between the electrodes by a "stepped leader", even though the electric field strength is less than the field strength for breakdown by Paschen's law. What one has here is a growing electrostatic instability, triggered by a small initial electric field inhomogeneity.

A preferred point for the beginning of a stepped leader is the field inhomogeneity near the triple point where the conductor, the gas and the insulator, meet.

### 3. Stabilization in a Drag-Free Taylor flow

It is known, and used in power switches, that a gas jet under high pressure can blow out an electric arc discharge. Recognizing the breakdown below the Paschen limit as a growing electrostatic instability, it is conjectured that much higher voltages can be reached if the onset of this instability is suppressed by a gas flow, with the stagnation pressure of the flow exceeding the electric pressure in between the electrodes with a large potential difference, overwhelming the electric pressure of a developing electric field inhomogeneity. It is for this reason proposed to levitate a spherical conductor by both hydrodynamic and magnetic forces inside a Taylor flow (Taylor, 1922), a special drag-free spiral flow [see Fig.1].

With the absence of drag forces on the sphere inside the horizontally flowing spiral Taylor flow, the sphere must only be levitated in the vertical direction by an externally applied magnetic field. By its levitation, the triple point as the source of a field inhomogeneity is thereby eliminated.

If for air, at a pressure of 1 atmosphere, the breakdown voltage is $3 \times 10^6$ V/m, it would (according to Paschen's law) for a pressure of 300 atmospheres be equal to $\sim 10^9$ V/m, that is for meter-size distances of the order $\sim 10^9$ Volt.

The electric field of $10^9$ V/cm is about $E \cong 3 \times 10^4$ [esu], with an electric pressure $E^2/8\pi \sim 4 \times 10^7 \, dyn/cm^2 \sim 40$ atmospheres. At a pressure of 300 atmospheres, air (or some other gas) would have a density of the order $\rho \sim$ 1g/cm$^3$. For the stagnation pressure p = (1/2) $\rho v^2$ of the Taylor flow moving with the velocity v [cm/s], to exceed the electric pressure, it is required that $(1/2)\rho v^2 > E^2/8\pi$, from which one obtains v ~ 100 meter/second. For a ten times smaller velocity one could reach $10^8$ Volt, with the other parameters remaining the same. Instead of a gas under high pressure one may also use a nonconducting fluid under normal pressure.

What remains is how to charge the sphere to such a high potential. There replacing the ribbon with seem to be two possibilities:

1. As in a Van de Graaff generator, by letting a charged nonconducting ribbon pass through the center of the sphere, releasing the charge in its center, or by replacing the ribbon with a stream of (positively) charged pellets.
2. By inductive charging in a rising magnetic field, releasing charges from the center of the sphere to the fluid flowing through a hole in the sphere.

### 4. Levitation in a Taylor Flow

The flow discovered by Taylor (Taylor, 1922) is the superposition of a uniform axial flow of constant velocity U with a constant swirl $W = (U/l)\,r$, where $r/l$ is a measure for the intensity of the swirl. In cylindrical coordinates the stream function $\psi$ (r, z) of the Taylor flow satisfies the equation (Squire, 1956):

$$\frac{\partial^2 \psi}{\partial z^2} + \frac{\partial^2 \psi}{\partial r^2} - \frac{1}{r}\frac{\partial \psi}{\partial r} + \frac{4}{l^2}\psi = \frac{2U}{l^2}r^2 \qquad (6)$$

For a different problem the solution of (6), in terms of Bessel and Neumann functions, has been given ($\kappa_1$, $\kappa_2$ constants of integration) by Moore and Leibovich (Moore and Leibovich, 1971):

$$\psi = \frac{1}{2}Ur^2\left(1 + \kappa_1 \frac{J_{3/2}(\xi)}{\xi^{3/2}} + \kappa_2 \frac{N_{3/2}(\xi)}{\xi^{3/2}}\right) \qquad (7)$$

where $\xi = \frac{2}{l}\sqrt{r^2 + z^2}$

and $J_{3/2}(\xi) = \sqrt{\frac{2}{\pi\xi}}\left(\frac{1}{\xi}\sin\xi - \cos\xi\right)$, $N_{3/2}(\xi) = \sqrt{\frac{2}{\pi\xi}}\left(\sin\xi + \frac{1}{\xi}\cos\xi\right)$.

The velocity components in the z, r and $\varphi$ direction are

$$\frac{u}{U} = 1 + \kappa_1 \left( \frac{J_{3/2}}{\xi^{3/2}} - 2\frac{r^2}{l^2} \frac{J_{5/2}}{\xi^{5/2}} \right) + \ldots \tag{8}$$

$$\frac{v}{U} = 2\kappa_1 \frac{zr}{l^2} \frac{J_{5/2}}{\xi^{5/2}} + \ldots \tag{9}$$

$$\frac{w}{U} = \frac{r}{l}\left(1 + \kappa_1 \frac{J_{3/2}}{\xi^{3/2}}\right) + \ldots \tag{10}$$

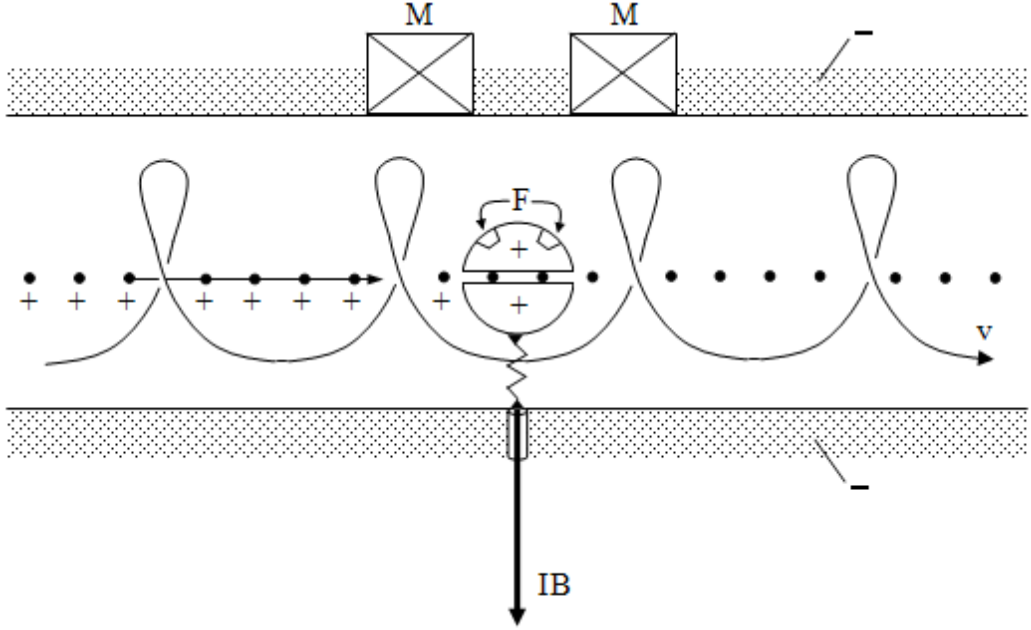

Fig.1. In drag-free Taylor flow magnetically levitated sphere to be charged up to ultrahigh potential by electrically charged pellets passing through the center of the sphere, **M** magnets, **F** ferromagnets, **IB** ion beam.

where the terms involving the Neumann functions and the second constant of integration $\kappa_2$ are represented by dots. The J-solution of (7), putting $\kappa_2 = 0$, has no singularity at $\xi = 0$, and is the solution of interest. With the appropriate boundary condition it is:

$$\psi = \frac{1}{2}Ur^2\left[1 - \left(\frac{2R/l}{\xi}\right)^{3/2} \frac{J_{3/2}(\xi)}{J_{3/2}(2R/l)}\right] \tag{11}$$

Now, if $J_{5/2}$ vanishes on the surface of a sphere placed in the Taylor flow, then all velocity components vanish on the surface of the sphere. And if $J_{5/2}$ is zero on the surface of the sphere, the circumferential shear vanishes as well. As a consequence, there is no boundary layer on the surface of the sphere and with it no drag. The pressure on the sphere is constant and the sphere stays at rest, except that it still is subject to a downward directed gravitational force, if U is directed horizontally. The downward force can be compensated by an externally applied magnetic field, making parts of the sphere from a ferromagnetic material.

We may add that in a beautiful experiment (Harvey, 1962) had verified the Taylor solution.

## 5. Discharging the Sphere

One simple way the highly charged sphere can be discharged is over a spark gap, to be formed by moving the sphere towards the wall with the help of the magnetic field holding the sphere in the center of the Taylor flow, until the pd product becomes smaller than the value where breakdown sets in below the Paschen curve. If the sphere is positively charged, and if the discharge current is larger than the Alfven current for electrons, this will favor a discharge into an intense ion beam below the Alfven limit, not only suitable as a driver for inertial confinement fusion, but for many other applications as well.

## Conclusion

The central problem for the achievement of the release of thermonuclear energy by inertial confinement is the energy and power flux density of the "driver", the means for igniting a thermonuclear reaction. These demands can be easily met by a fusion explosive, but ignition should still be easy with 10-100 MJ driver energy and a power in excess of $10^{15}$ Watt, to be focused onto an area smaller than 1 cm$^2$. This is much less than the energy and power released by a fission explosion, but nevertheless very difficult to reach. At the present, great efforts under way to realize drivers with an energy of just one megajoule. It appears that ignition might there marginally become possible. The limitation in available driver energy reduces the ignition to a problem of miniaturization, where all the possible instabilities, like the Rayleigh Taylor instability, become important. These problems could altogether largely be avoided with 10 to 100 times larger driver energies. And with 100 times larger driver energies the ignition of pure deuterium fusion explosions should become possible.

When back in 1968 I had suggested to use Marx generators for thermonuclear ignition, my reasoning was that with higher voltages higher power could be achieved. However, to reach gigavolt potentials much more radical ideas must be explored.